\documentclass[twocolumn,superscriptaddress,aps,prl,showpacs,preprintnumbers,amsmath,amssymb,floatfix]{revtex4-1}

\usepackage[latin9]{inputenc}
\usepackage{color}
\usepackage{graphicx}
\usepackage{esint}
\usepackage[unicode=true,
 bookmarks=false,
 breaklinks=false,pdfborder={0 0 1},backref=false,colorlinks=true]
 {hyperref}

\makeatletter
\@ifundefined{textcolor}{}
{%
 \definecolor{BLACK}{gray}{0}
 \definecolor{WHITE}{gray}{1}
 \definecolor{RED}{rgb}{1,0,0}
 \definecolor{GREEN}{rgb}{0,1,0}
 \definecolor{BLUE}{rgb}{0,0,1}
 \definecolor{CYAN}{cmyk}{1,0,0,0}
 \definecolor{MAGENTA}{cmyk}{0,1,0,0}
 \definecolor{YELLOW}{cmyk}{0,0,1,0}
}

\usepackage{dcolumn}
\usepackage{bm}
\usepackage{color}
\usepackage{braket}

\newcommand{\bD}{{\cal D}}
\newcommand{\B}{{\cal B}}

\def\e{\mathcal{E}}
\def\ba{\begin{eqnarray}}
\def\ea{\end{eqnarray}}
\def\beq{\begin{equation}}
\def\eeq{\end{equation}}

\usepackage{braket}

\def\d{\downarrow}
\def\u{\uparrow}
\def\D{\Downarrow}
\def\U{\Uparrow}

\def\e{\mathcal{E}}

\makeatother

\begin{document}

\title{Fractional Quantum Hall States of Rydberg Polaritons}

\author{Mohammad F. Maghrebi}
\affiliation{Joint Quantum Institute, NIST/University of Maryland, College Park, Maryland 20742, USA}

\author{Norman Y. Yao}
\affiliation{Department of Physics, University of California, Berkeley, California 94720, USA}

\author{Mohammad Hafezi}
\affiliation{Joint Quantum Institute, NIST/University of Maryland, College Park, Maryland 20742, USA}
\affiliation{Department of Electrical Engineering and Institute for Research in Electronics and Applied Physics,
University of Maryland, College Park, MD 20742, USA}

\author{Thomas Pohl}
\affiliation{Max Planck Institute for the Physics of Complex Systems, 01187 Dresden, Germany}

\author{Ofer Firstenberg}
\affiliation{Department of Physics of Complex Systems, Weizmann Institute of Science, Rehovot 76100, Israel}

\author{Alexey V. Gorshkov}
\affiliation{Joint Quantum Institute, NIST/University of Maryland, College Park, Maryland 20742, USA}

\date{\today}
\begin{abstract}
We propose a scheme for realizing fractional quantum Hall states of light. In our scheme, photons of two polarizations are coupled to different atomic Rydberg states to form two flavors of Rydberg polaritons that behave as an effective spin. An array of optical cavity modes overlapping with the atomic cloud enables the realization of an effective spin-$1/2$ lattice. We show that the dipolar interaction between such polaritons, inherited from the Rydberg states, can be exploited to create a flat, topological band  for a single spin-flip excitation. At half filling, this gives rise to a photonic (or polaritonic) fractional Chern insulator -- a lattice-based, fractional quantum Hall state of light. 
\end{abstract}


\maketitle

Fractional Chern insulators are exotic topological phases of matter that can be thought of as magnetic-field-free fractional quantum Hall states on a lattice 
 \cite{regnault11}.
Recently, there have been several proposals to implement fractional Chern insulators in optical flux lattices \cite{cooper13b} and dipolar systems \cite{yao13c}. 
On the other hand, the recent experimental realization of topological band structures in arrays of photonic modes \cite{hafezi13,rechtsman13} points to the intriguing possibility of realizing strongly-correlated interacting topological states of light \cite{hafezi13c,umucalilar13}. Given that  photonic systems are prepared and probed differently \cite{aspuru-guzik12}, typically have no chemical potential \cite{hafezi14}, and
 exhibit different decoherence mechanisms \cite{kapit14,grusdt14}
as compared to their electronic counterparts, interacting topological states of light will open new avenues to the study of exotic physics \cite{hafezi13c}. Furthermore,  such states might  enable the construction of numerous robust, i.e.\ topologically protected, optical devices such as filters \cite{hafezi13c}, switches, and delay lines \cite{hafezi11b,yao13b}. Finally, once such highly non-classical states of light are released onto freely propagating non-interacting modes, they might be usable as resources for enhanced precision measurements and imaging \cite{lloyd08b}.

While strong interactions between microwave photons  are readily achievable \cite{Schoelkopf08, Houck12,Devoret13,Koch10, Nunnenkamp13,Kapit13}, the realization of strong high-fidelity interactions between optical photons has remained a challenge \cite{birnbaum05,fushman08,reiserer14}. 
Only recently, the required strong interaction between optical photons has been implemented in a robust fashion by transforming photons into superpositions of light and highly excited atomic Rydberg states, thus forming polaritons. These polaritons  inherit strong dipolar interactions from Rydberg states \cite{friedler05,gorshkov11,dudin12,parigi12,firstenberg13,maxwell13,gunter13,baur14,gorniaczyk14} and -- together with artificial gauge fields that arise naturally in dipolar systems via the Einstein-de-Haas effect \cite{yao12d,yao13c,kiffner13} -- constitute an ideal platform for realizing interacting topological states of light \cite{cho08b,nielsen10b,grusdt13,umucalilar12}. 

\begin{figure}[b!]
\begin{center}
\includegraphics[width = 0.98 \columnwidth]{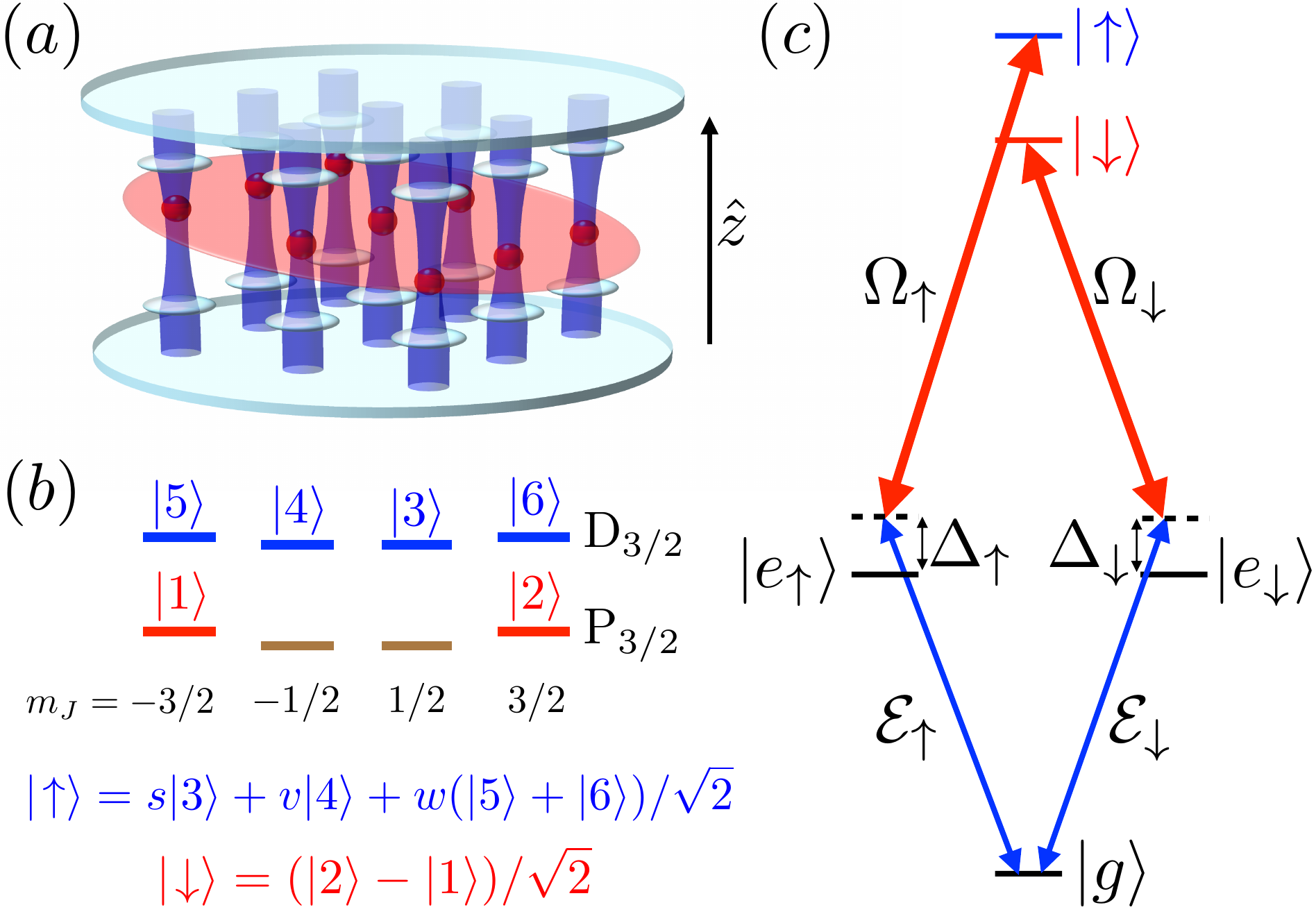}
\caption{(a) A quasi-two-dimensional cloud of atoms (red disk)
overlaps with an array of cavity modes (with cavity axis $\hat z$) at a plane tilted relative to $\hat z$. The overlaps (red balls) allow one to define a square-lattice array
of Rydberg polaritons. Each polariton can be in state $\ket{\Uparrow}$ or $\ket{\Downarrow}$. The resulting spin model has a fractional quantum Hall ground state. (b) To achieve a topological flat-band structure, single-atom dressed states $\ket{\u}$ and $\ket{\d}$ are constructed as linear combinations of several Rydberg levels with spatially dependent coefficients $s$, $v$, and $w$. 
A weak DC electric field along $\hat z$ is assumed. (c) The $\ket{\Uparrow}$ and $\ket{\Downarrow}$ polaritons are created by coupling $\ket{\u}$ and $\ket{\d}$ 
states to $\e_\u$ ($\sigma^-$-polarized) and $\e_\d$ ($\sigma^+$-polarized) photonic modes, respectively. The flip-flop ($\ket{\u \d} \rightarrow \ket{\d \u}$) dipolar interaction 
yields the fractional quantum Hall polariton Hamiltonian.  \label{fig:scheme}}
\end{center}
\end{figure}

In this Letter, we present the first example of a fractional Chern insulator of photons (or polaritons) in such a medium. The particular insulator we construct corresponds to the $\nu=1/2$ filling fraction of the familiar Laughlin fractional quantum Hall state, in which an additional injected polariton fractionalizes into a pair of quasiparticles obeying semionic statistics. In the absence of quantized light, our proposal also allows one to implement fractional Chern insulators of Rydberg atoms. 

To understand the basic idea [see Fig.\ \ref{fig:scheme}(a)], consider a cloud of atoms whose overlaps with spatially separated  optical modes  
form a square lattice (notice that, conveniently, we do not require an array of traps). 
Under the conditions of electromagnetically induced transparency (EIT) \cite{fleischhauer05}, an auxiliary control field coherently couples a $\sigma^+$-polarized photon to a Rydberg state creating a hybrid atom-photon excitation called a Rydberg polariton (call it the $\D$-polariton) \cite{guerlin10,brion12,parigi12,zhang13b,grankin14}. Another control field couples a $\sigma^-$-polarized photon to a different Rydberg state creating a different polariton, the $\U$-polariton. We populate each site  of the lattice (i.e.\ each optical mode)  
with exactly one polariton (either $\U$ or $\D$), so that each  
site becomes an effective spin-$1/2$ particle. Rydberg polaritons inherit dipolar interactions between Rydberg atoms giving rise to long-range flip-flop interactions between polaritonic spins. Thinking of the Rydberg-polariton spin-flip operator $\ket{\U} \bra{\D}$ on one site as a bosonic creation operator, we obtain a model of hardcore bosons hopping on a square lattice. Applied electromagnetic fields can be used to break time-reversal symmetry and to tune Rydberg-Rydberg dipolar interactions into creating a topological flat band for these bosons. In particular, the resulting complex-valued hopping amplitudes endow the bands with non-trivial topology,  characterized by a nonzero Chern number of the bands. 
By analogy with Landau levels, the flatness of the topological band manifests itself in the smallness of the band's dispersion relative to the bandgap and allows hardcore interactions to turn a fractionally filled topological band into a fractional Chern insulator. In the language of spins, the $\nu = 1/2$ Laughlin state that we obtain is 
a gapped chiral spin liquid \cite{kalmeyer87,kalmeyer89}. A limiting case of this proposal corresponds to polaritons consisting entirely of Rydberg excitations, which gives rise to a fractional Chern insulator of Rydberg atoms even in the case where each effective spin consists of a single  atom.

\textit{Engineering the polariton Hamiltonian.}---As shown in Fig.\ \ref{fig:scheme}(a), the atomic cloud 
is trapped in a plane tilted relative to the cavity axis $\hat z$. As we show below, this variable tilt is crucial for achieving  sufficient controllability to obtain flat topological bands.

As shown in Fig.\ \ref{fig:scheme}(b), a DC electric field is applied along $\hat z$ to remove the degeneracy between Zeeman levels with different $|m_J|$ for Rydberg states P$_{3/2}$ and D$_{3/2}$  (principal quantum numbers will be specified below).  At the same time, the field is assumed to be sufficiently weak that the induced dipole moments are negligible.  
Auxiliary optical and microwave fields can be used to define dressed states $\ket{\d} = (\ket{2} - \ket{1})/\sqrt{2}$ and $\ket{\u} = s \ket{3} + v \ket{4} + w (\ket{5}+\ket{6})/\sqrt{2}$, where the complex coefficients $s$, $v$, and $w$ vary from site to site. 
The dipolar flip-flop interaction takes two Rydberg atoms in state $\ket{\textrm{P}_{3/2}}\ket{\textrm{D}_{3/2}}$ and produces the state $\ket{\textrm{D}_{3/2}}\ket{\textrm{P}_{3/2}}$. Projecting this interaction onto states $\ket{\u}$ and $\ket{\d}$, we obtain
\begin{equation} 
  H_\textrm{I} =\sum_{A,B,i\in A, \\ j\in B } t_{ij} 
  \, \sigma_{\uparrow \downarrow}^i  \sigma_{\downarrow \uparrow}^j, \label{eq:updown}
\end{equation}
where $\sigma^i_{\alpha \beta} = \ket{\alpha}_i \bra{\beta}_i$, $A$ and $B$ label the sites of the array, $i$ and $j$ run over the atoms on sites $A$ and $B$, respectively. The amplitudes $t_{ij}$ can be tuned by adjusting the site-dependence of $s$, $v$, and $w$ and by adjusting the direction of the $\hat z$ axis relative to    
the spin lattice.

We will now use Eq.\ (\ref{eq:updown}) to derive an interaction between Rydberg polaritons. We start with an ensemble of 
effective five-level atoms  
on each site of the square lattice [see Fig.\ \ref{fig:scheme}(c)]: the ground state $\ket{g}$,  excited states $\ket{e_\u}$ and $\ket{e_\d}$, and the dressed Rydberg states $\ket{\u}$ and $\ket{\d}$ defined above. 
Since most 
atoms will remain in 
state $\ket{g}$, atomic excitations can be described using bosonic operators acting on state $|g \dots g\rangle$ 
\cite{fleischhauer00,gorshkov07b}. 
We take $g_{\alpha j} = g_\alpha \sin(\omega_{1, \alpha} z_j/c)$  to be the coupling constant between atom $j$ at position $z_j$ and optical mode  
$A$ of frequency $\omega_{1,\alpha}$,  polarization $\alpha = \u, \d$, and with creation operator $a^\dagger_{A,\alpha}$. 
We further assume that the two-photon-resonant running-wave control fields of frequency $\omega_{2,\alpha}$ and Rabi frequency $\Omega_\alpha$ are propagating along $\hat z$. We can then define the following -- slowly-varying in time -- collective operators for  
site $A$ \cite{fleischhauer00,gorshkov07b}.
$\e^\dagger_{A,\alpha} = a^\dagger_{A,\alpha} e^{- i \omega_{1,\alpha} t}$  creates a photon of polarization $\alpha$,  $P^\dagger_{A,\alpha} = (1/g^{\textrm{col}}_{A,\alpha}) \sum_{j \in A} g_{\alpha j}  \sigma^{j}_{e_\alpha,g} e^{- i \omega_{1,\alpha} t}$
creates a collective $\ket{e_\alpha}$ excitation, while 
$S^\dagger_{A,\alpha} = (1/g^\textrm{col}_{A,\alpha}) \sum_{j \in A} g_{\alpha j}  \sigma^{j}_{\alpha,g} e^{- i (\omega_{1, \alpha}+\omega_{2, \alpha})t+i\omega_{2,\alpha}z_j/c}$
creates a collective $\ket{\alpha}$ excitation. 
Here $g^\textrm{col}_{A,\alpha} = \sqrt{\sum_{j \in A} |g_{\alpha j}|^2}$ is the collectively enhanced atom-photon coupling. This collective enhancement is the main reason for using entire atomic ensembles in place of single atoms as this allows one to achieve strong coupling even when individual atoms are coupled to optical modes  
weakly.   The non-interacting Hamiltonian in the rotating frame becomes
\ba
 H_0&=& \sum_{A,  \alpha}- (\Delta_\alpha+i\gamma)\,P_{A,\alpha}^\dagger P_{A,\alpha} + \Big(g^\textrm{col}_{A,\alpha} P_{A,\alpha}^\dagger \e_{A,\alpha} \nonumber \\ 
 &&+ \Omega_\alpha S_{A,\alpha}^\dagger P_{A,\alpha}+ {\rm h.c.}\Big), 
\ea 
where $\Delta_\alpha$ is the single-photon detuning, and $2 \gamma$ is the decay rate of $\ket{e_\alpha}$. 
The Hamiltonian can be diagonalized in the dark and bright polariton basis,
$
  H_0=\sum_{A,\alpha} E_{A,\B_{1\alpha}} \B^\dagger_{A,1\alpha} \B_{A,1\alpha}+E_{A,\B_{2\alpha }} \B^\dagger_{A,2\alpha} \B_{A,2\alpha}.
$
The dark polariton  $\bD_{A,\alpha}^\dagger = (g^\textrm{col}_{A,\alpha} S_{A,\alpha}^\dagger - \Omega_\alpha \e_{A,\alpha}^\dagger)/\sqrt{|g^\textrm{col}_{A,\alpha}|^2 + \Omega_\alpha^2}$ has zero energy (in the rotating frame)  and thus does not appear in $H_0$, 
while the two 
bright polaritons (linear combinations of $\Omega_\alpha S_{A,\alpha}^\dagger + g^\textrm{col}_{A,\alpha} \e_{A,\alpha}^\dagger$ and $P^\dagger_{A,\alpha}$)  have large energies (with imaginary parts due to the decay rate $2 \gamma$ of $\ket{e_\alpha}$). 
Provided that this energy is larger than the strength of the Rydberg interaction between sites, this interaction will be too weak to convert dark polaritons into bright ones, ensuring that the total number of  dark polaritons is conserved. Therefore, we can consider the subspace consisting solely of dark polaritons, for which $H_0 = 0$. 

The interaction between dark polaritons is mediated via a long-range exchange interaction at different sites, as provided in Eq.\ (\ref{eq:updown}). The indices $i$ and $j$ belong to different sites since we assume that the system starts with one Rydberg excitation per site and since time evolution will not change this. Indeed, a hopping of a Rydberg excitation from one site onto another requires a flip-flop on an optical transition, which will be negligible for our intersite separations. The hopping will be further suppressed by  interactions between two Rydberg excitations on the same site. 
Therefore, 
using $\sigma_{\uparrow \downarrow}^i =\sigma_{\uparrow g}^i \, \sigma_{g \downarrow}^i$, 
the interacting Hamiltonian becomes
\begin{equation}\label{Eq: Hamiltonian in the NR frame}
  H=\sum_{A\ne B}{t_{AB}} \sum_{i\in A}\sigma_{\uparrow g}^i \, \sigma_{g \downarrow}^i \sum_{ j\in B } \sigma_{\downarrow g}^j \, \sigma_{g\uparrow}^j.
\end{equation}
Note that the interaction amplitude $t_{AB}$ depends only on the site index (and not specific atoms within each site) as the distance between two sites is much greater than the distribution size of atoms in a single one, in analogy with Ref. \cite{weimer13b}.

The next step is to rewrite the Hamiltonian in terms of collective operators $S^\dagger_{A,\alpha}$ in place of the microscopic atomic operators $\sigma^j_{\alpha,g}$.  
 For our parameters, for any $j \in A$, $[\sin(\omega_{1, \u} z_j/c)/\sin(\omega_{1, \d} z_j/c)] \exp[i (\omega_{2,\u}-\omega_{2,\d})z_j/c] \approx 1$ up to an $A$-dependent phase, which can be absorbed in the definition of $S^\dagger_{A,\alpha}$ \cite{supp}.  
One can then check that the Hilbert space spanned on each site by $S^\dagger_{A,\u}|g\cdots g\rangle_A$ and $S^\dagger_{A,\d}|g\cdots g\rangle_A$ is closed under the action of the Hamiltonian (\ref{Eq: Hamiltonian in the NR frame}), which allows us to rewrite Eq.\ (\ref{Eq: Hamiltonian in the NR frame}) within this Hilbert space, in the rotating frame, as
\begin{equation}
  H=\sum_{A \neq B} {t_{AB}} S^\dagger_{A,\uparrow} S_{A,\downarrow} S^\dagger_{B,\downarrow} S_{B,\uparrow}.  \label{eq:Seq}
\end{equation}

We now recall that we are restricted to a subspace consisting of dark polaritons, $\ket{\Uparrow}_A = \bD^\dagger_{A,\uparrow}|g\cdots g\rangle$ 
and  $\ket{\Downarrow}_A = \bD^\dagger_{A,\downarrow}|g\cdots g\rangle$. 
Since $g_{A,\alpha}^{\rm col}\gg \Omega_\alpha$, the dark polaritons are predominantly composed of Rydberg excitations. The atomic interactions, therefore, directly map onto polariton interactions, irrespective of the precise value of $g_{A,\alpha}^{\rm col}$ (which can depend on $A$ due to atom-number variations). 
Consequently, we are arrive at the  final polariton Hamiltonian
\begin{equation}
  H=\sum_{A \neq B} {t_{AB}} \, \bD^\dagger_{A,\uparrow} \bD_{A,\downarrow} \bD^\dagger_{B,\downarrow} \bD_{B,\uparrow}, \label{eq:final}
\end{equation}
which will be used to realize a topological flat band and a fractional Chern insulator by tuning the site-dependent interaction $t_{AB}$.

\begin{figure}[b!]
\begin{center}
\includegraphics[width = 0.98 \columnwidth]{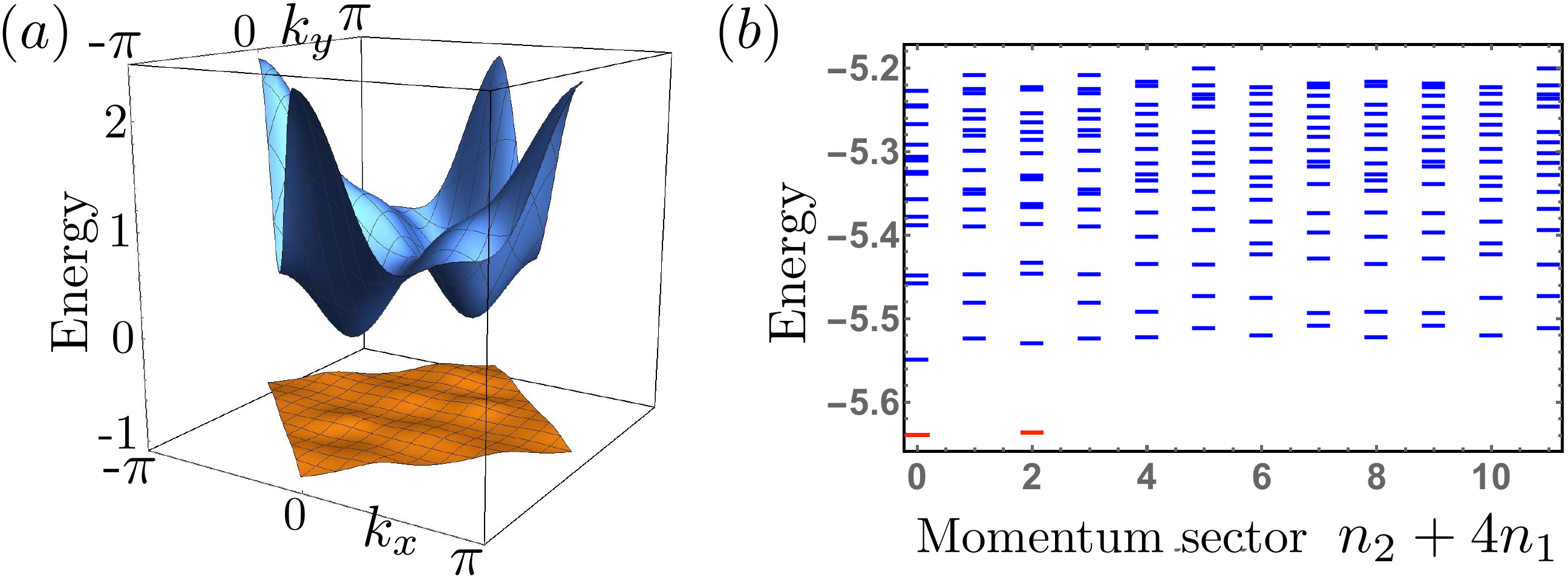}
\caption{(a) Topological flat band for Rydberg polaritons (Chern number $C = -1$) featuring a flatness (band gap divided by band width) $\approx 10$.  (b) Fractional Chern insulator of Rydberg polaritons. For a $6 \times 4$ lattice with $6$ particles with periodic boundary conditions, 
the eigenstates in momentum sector $n_2 + 4 n_1$, where $(k_x,k_y) = (n_1/3,n_2/2-n_1/3) \pi$, $n_1 = 0,1,2$, and $n_2 = 0,1,2,3$. 
The two degenerate ground states (red) at $(k_x,k_y) = (0,0)$ and $(0,\pi)$ are separated from the other states by a gap.
   \label{fig:FCI}}
\end{center}
\end{figure}

\textit{Fractional quantum Hall states of Rydberg polaritons.}---Thinking of $\ket{\Downarrow}$ as vacuum and $\ket{\Uparrow}$ as the presence of a hardcore boson, Eq.\ (\ref{eq:final}) describes the hopping of such hardcore bosons. Following a recipe similar to Ref.\ \cite{yao13c}, the site-dependent parameters $s$, $v$, and $w$ are chosen to yield a lattice with a two-site unit cell. These parameters, together with the direction of the quantization axis $\hat z$ relative to the  
spin lattice, are then tuned \cite{supp} to achieve a topological flat band for the bosons, as shown in Fig.\ \ref{fig:FCI}(a). The band's flatness (ratio of band gap to band width) is $\approx 10$, while its Chern number is $C = -1$, meaning that the band is topological.

We now consider filling the band with bosons to a filling $\nu = 1/2$, i.e.\ half a boson per unit cell. To show that the hardcore interactions alone suffice to produce a fractional Chern insulator, we diagonalized the Hamiltonian (\ref{eq:final}) on a $6 \times 4$ lattice with periodic boundary conditions (we also verified that our results hold for an $8 \times 4$ lattice). As shown in Fig.\ \ref{fig:FCI}(b), we obtain the two-fold degenerate ground state separated from the rest of the eigenstates by a gap, consistent with the $\nu = 1/2$ Laughlin state on a torus \cite{wang11e}. As an additional diagnostic, we compute the many-body Chern number. To do this, we numerically calculate the ground-state wavefunction $\ket{\Psi}$ in the presence of boundary-condition twists $(\theta_x,\theta_y)$, which are equivalent to inserting fluxes \cite{regnault11}. The many-body Chern number, which is analogous to the Hall conductance, is then defined as $\sigma_{xy} = \frac{1}{2 \pi} \int \int F(\theta_x,\theta_y) d \theta_x d \theta_y$, where the many-body Berry curvature is $F(\theta_x,\theta_y) = \textrm{Im}\left(\langle \partial_{\theta_y} \Psi|\partial_{\theta_x} \Psi\rangle - \langle \partial_{\theta_x} \Psi|\partial_{\theta_y} \Psi\rangle\right)$. For both of our degenerate ground states, we find $\sigma_{xy} = - 0.5$, consistent with the $\nu = 1/2$ Laughlin state, or equivalently the Kalmeyer-Laughlin chiral spin liquid \cite{kalmeyer87,kalmeyer89}.

\textit{Experimental considerations.}---Let us begin by emphasizing that in the simplest case where the polaritons have a vanishing photonic component, we obtain a spin model Eq.\ (\ref{eq:Seq}), where each spin state is a collective Rydberg excitation. The implementation of this purely atomic spin model is a natural intermediate step towards the realization of the polaritonic fractional Chern insulator, a step that can make use of  
cavity modes for addressing  individual collective spins. Such a  purely atomic implementation also works with a single atom per site, in which case Eq.\ (\ref{eq:updown}) immediately yields the desired spin Hamiltonian.  Given the strength of Rydberg interactions, this is a promising implementation of fractional Chern insulators in optical lattices \cite{schauss12,zhang11c} or microtrap arrays \cite{piotrowicz13,nogrette14}.

 The array of cavities can be created using arrays of microlenses \cite{ErtmerPRL2002} or spherical micromirrors \cite{LoktevSPIE2003}. For example, as shown in Fig.\ \ref{fig:scheme}(a), two microlens arrays enclosed in a cavity  
with planar mirrors 
can support an array of 
Gaussian modes. Using $69$D$_{3/2}$ and $70$P$_{3/2}$ of ${}^{87}$Rb for $\ket{\u}$ and $\ket{\d}$, an $85$ $\mu$m lattice constant gives nearest-neighbor dipole-dipole interactions $V_\textrm{dd}/2\pi = 60$ kHz, larger than a reasonable cavity decay rate $\kappa/2\pi \sim 10$ kHz \cite{notcutt05} and Rydberg decay rate $\lesssim (2\pi) 1$  kHz. The waists of cavity modes, which define the polaritons, 
are taken to be $< 10$ $\mu$m, the blockade radius of our Rydberg states. The control fields $\Omega_{\u,\d}$ can be spatially uniform and address all sites globally. The auxiliary optical fields (Rabi frequency $\Omega_\textrm{dr}$) used to create the dressed state $\ket{\u}$ are uniform over each site but differ between sites in the checkerboard  fashion necessary to create the desired fractional Chern insulator \cite{supp}. 
We then 
adopt the following 
ladder of energy scales: $(\Omega_\textrm{dr}/2 \pi = 10 \textrm{ MHz}) \gg (\Omega_{\u,\d}/2 \pi = 2  \textrm{ MHz}) \gg (\omega_\textrm{EIT}/2 \pi = 300 \textrm{ kHz}) \gg (V_\textrm{dd}/2 \pi = 60 \textrm{ kHz})$. The condition $\Omega_\textrm{dr} \gg \Omega_{\u,\d}$ ensures that the control fields couple to the dark states $\ket{\u}$ and $\ket{\d}$ 
but not to the bright states.  The condition $\Omega_{\u,\d} \gg \omega_\textrm{EIT}$, with $\omega_\textrm{EIT} \sim \Omega_{\u,\d}^2/|\Delta_{\u,\d}|$ the EIT linewidth \cite{fleischhauer05}, arises from the requirement $|\Delta_{\u}-\Delta_{\d}| \gg \Omega_{\d}$, which prevents two-photon resonant coupling of $\e_\u$ to $\Omega_\d$ \cite{supp}. The condition $\omega_\textrm{EIT} \gg V_\textrm{dd}$ ensures that interactions  
do not violate EIT.  

The preparation of the fractional Chern insulator state can be achieved as follows.  By changing the direction and strength of the applied DC electric field, one first tunes the interaction Hamiltonian to the part of the phase diagram where the ground state is a simple solid or superfluid, 
in which each effective spin is in a well-defined state. 
One then prepares the atomic state corresponding to this solid or superfluid by introducing an appropriate single collective Rydberg excitation  
onto each lattice site \cite{gorshkov13b}. By analogy with  
Ref.\ \cite{dudin12}, this can be done by relying on Rydberg blockade and driving a two-photon transition to the Rydberg state, where the bottom leg of the transition uses the 
cavity mode. 
The variations in the collective Rabi frequency $g^\textrm{col}_{A,\alpha}$ from one lattice site  
to another can be mitigated by
using adiabatic preparation \cite{pohl10,schaus14}. One then changes the parameters of the Hamiltonian (by tuning the DC electric field strength and direction) to adiabatically go across a phase transition (believed to be continuous \cite{barkeshli14b}) into the fractional Chern insulator phase of collective Rydberg excitations. Finally, the control fields are turned on to adiabatically convert the Rydberg excitations $S^\dagger_\downarrow$ and $S^\dagger_\uparrow$ into polaritons $\ket{\Downarrow}$ and $\ket{\Uparrow}$, respectively.  
The addition of an auxiliary lattice of qubits with 
fast decay may  
provide an alternative elegantly preparation scheme and may further reduce the effects of photon loss   \cite{kapit14}. 

By analogy with Ref.\ \cite{yao13c}, one detection approach would attempt to flip the polaritons between $\ket{\Downarrow}$ and $\ket{\Uparrow}$ with a variable detuning and variable spatial dependence effectively realizing Bragg spectroscopy and providing the energy- and momentum-dependent spectral function. The spectral function can, in turn, be used to identify, for example, gapless chiral Luttinger liquids on the edge \cite{kjall12,goldman12b} and a spectral gap in the bulk. Another elegant detection approach, unique to the polaritonic fractional Chern insulators, relies on the retrieval \cite{fleischhauer00c,gorshkov07b} of the fractional Chern insulator state onto a purely photonic state of $\e^\dagger_\uparrow$ and $\e^\dagger_\downarrow$ photons. 
Classical and quantum correlations between the retrieved  photons could then be measured using 
quantum optics techniques  
and compared to those of the desired fractional Chern insulator \cite{hafezi13c}. 
Finally, an elegant combination of preparation and detection would involve first turning on all dressing and control fields to create a ``topological filter" \cite{hafezi13c}, and then sending single photons into each cavity; provided the incoming energy matches that of the fractional Chern insulator, the photonic fractional Chern insulator will be transmitted with probability determined by its (small) overlap with the input.

\textit{Outlook.}---While we have presented one of the most conceptually straightforward implementations of one of the simplest topological states, 
the ideas and methods presented in this Letter point to strongly interacting Rydberg polaritons as a very promising and powerful platform for realizing interacting topological states of light. In particular, it should be straightforward to extend to Rydberg polaritons dipolar-spin-model implementations \cite{dysp,peter14} of fractional Chern insulators in flat bands with arbitrary Chern numbers \cite{yang12,liu12c}. We also expect optical-flux-lattice approaches \cite{cooper13b} and approaches, in which the role of time is played by the propagation direction \cite{rechtsman13}, to be extendable to light. It is also natural to consider trapping ensembles of Rydberg atoms near arrays of optical-ring resonators in order to harness the recently demonstrated topological band structures in such systems \cite{hafezi13} for the creation of interacting topological states of light.

We used a separation of energy scales to provide a controllable way of creating  a 
long-range-entangled \cite{swingle14} topological state of  polaritons. At the same time, an experimentally more straightforward approach would consist of a free-space setup, in which spatially inhomogeneous control fields give rise to propagating polaritons. Our controllable creation of long-range-entangled topological states will motivate the study of this much more complex problem, in which topological and other exotic phenomena may manifest themselves in system dynamics.

We thank N.\ Henkel, M.\ Lukin, and Z.-X.\ Gong  for discussions. This work was supported by the NSF PFC at the JQI, NSF PIF, ARO, ARL, ARO MURI, AFOSR, the Miller Institute for Basic Research in Science, and the EU through the Marie Curie ITN ``COHERENCE" and EU-FET Grant No.\ HAIRS 612862.

\appendix

\renewcommand{\thesection}{S\arabic{section}} 
\renewcommand{\theequation}{S\arabic{equation}}
\renewcommand{\thefigure}{S\arabic{figure}}
\setcounter{equation}{0}
\setcounter{figure}{0}

\section*{Supplementary Online Material: Details of the experimental implementation in ${}^{87}$Rb}

 In this supplementary online material, we first present the details of the experimental implementation in ${}^{87}$Rb, including the construction of dressed states $\ket{\u}$ and $\ket{\d}$ in Fig.\ \ref{fig:scheme} of the main text.  We then present the dependence of hopping amplitudes $t_{AB}$ in Eq.\ (\ref{eq:final}) of the main text on dressing parameters $s$, $v$, and $w$, and give the values of these parameters that were used to construct Fig.\ \ref{fig:FCI} of the main text.

\begin{figure}[t!]
\begin{center}
\includegraphics[width = 0.98 \columnwidth]{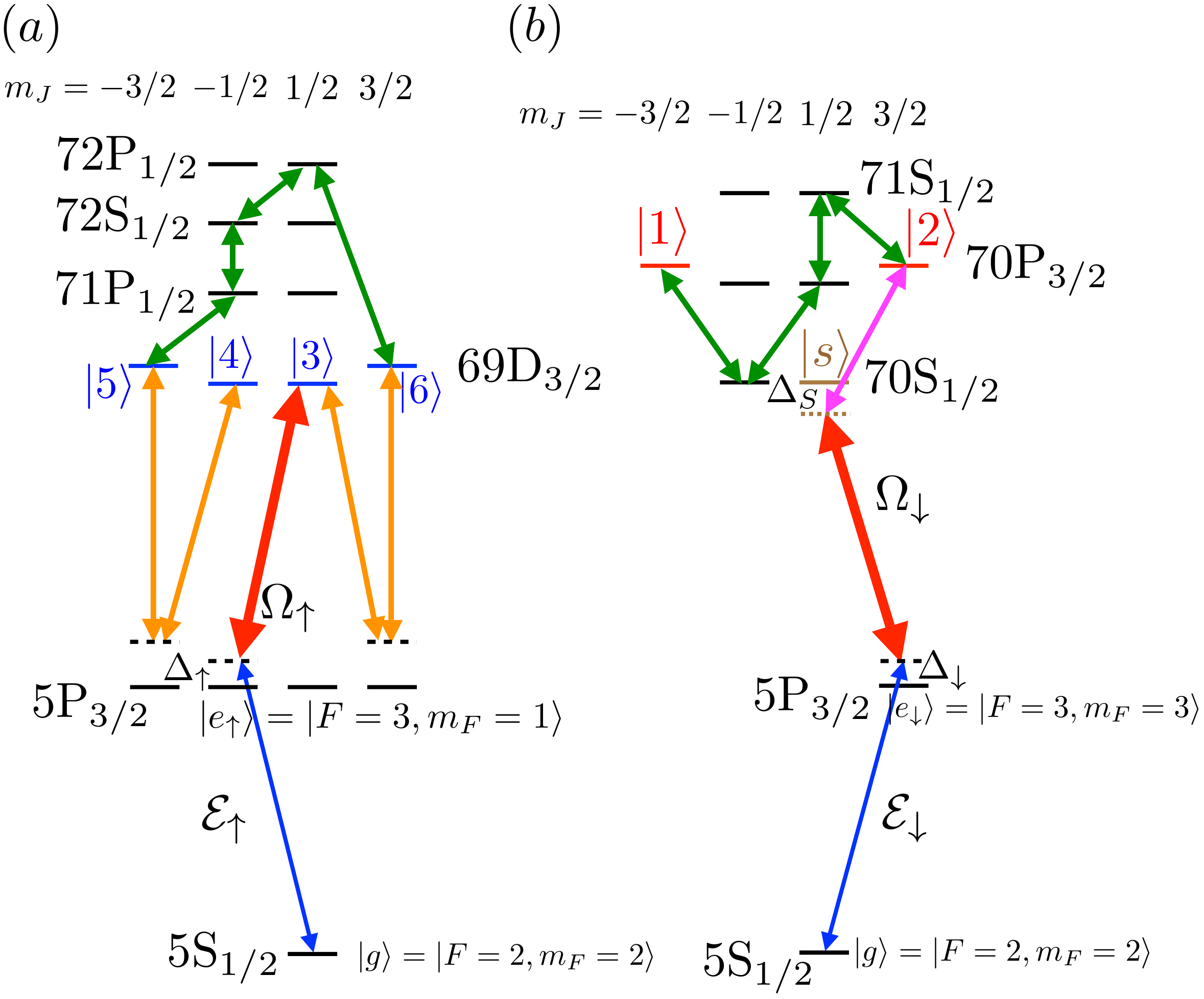}
\caption{Levels of ${}^{87}$Rb for generating the $\Uparrow$ (a) and $\Downarrow$ (b) polaritons. In particular, $\ket{\u} = s \ket{3} + v \ket{4} + w (\ket{5} + \ket{6})/\sqrt{2}$, while $\ket{\d} = (\ket{2} - \ket{1})/\sqrt{2}$. All the labeled states, except for $\ket{e_\u}$ have $m_I = 3/2$.  Green and magenta are microwave fields coupling Rydberg states to Rydberg states. Blue, red, and orange are optical fields coupling $5$P to $5$S and to Rydberg states.  $\ket{g} = \ket{F=2,m_F = 2} = \ket{m_J = 1/2, m_I = 3/2}$, $\ket{e_\d} = \ket{F = 3, m_F = 3} = \ket{m_J = 3/2, m_I = 3/2}$. The state $\ket{s}$ in (b) is used virtually.  The positions of Rydberg levels in (b) relative to those in (a) are drawn to minimize the crowding of the figure: in reality, $69$D$_{3/2}$ lies between $70$P$_{3/2}$ and $71$S$_{1/2}$.     \label{fig:real}}
\end{center}
\end{figure}
Expanding on Figs.\ \ref{fig:scheme}(b,c) in the main text, on the example of ${}^{87}$Rb, Fig.\ \ref{fig:real} shows a detailed level structure for constructing dressed states $\ket{\u}= s \ket{3} + v \ket{4} + w (\ket{5} + \ket{6})/\sqrt{2}$ and $\ket{\d}= (\ket{2} - \ket{1})/\sqrt{2}$ and for coupling these states to quantized light fields to form $\ket{\U}$ and $\ket{\D}$ Rydberg polaritons.  
By analogy with Ref.\ \cite{yao13c}, the optical Raman dressing beams coupling state $\ket{4}$ to state $\ket{5}$ and state $\ket{3}$ to state $\ket{6}$ provide the required spatial dependence (to be discussed below) of the dressing parameters $s$, $v$, and $w$. All the remaining classical fields (control fields $\Omega_\u$ and $\Omega_\d$, as well as microwave fields coupling $\ket{5}$ to $\ket{6}$, $\ket{1}$ to $\ket{2}$, and $\ket{2}$ to $\ket{s}$) are spatially uniform. State $\ket{\u}$ in Fig.\ \ref{fig:real}(a) (varying from 
site to site in a checkerboard fashion as discussed below) is the dark state of the Raman beams and the four microwave fields, while state $\ket{\d}$ (same on all sites) 
in Fig.\ \ref{fig:real}(b) is the dark state of the four microwave fields.  The $\ket{s} \rightarrow \ket{2}$ microwave field connects the odd-parity P-state $\ket{\d}$ to the even-parity state $70$S$_{1/2}$, where the latter state is assumed to be detuned ($\Delta_S \gg \Delta_\d$) and is used virtually.

Let us now show explicitly how the Raman dressing beams coupling state $\ket{5}$ to state $\ket{4}$ and state $\ket{3}$ to state $\ket{6}$, together with the microwave dressing fields coupling state $\ket{5}$ to state $\ket{6}$, can turn $\ket{\u} = s \ket{3} + v \ket{4} + w (\ket{5} + \ket{6})/\sqrt{2}$ into the dark state (for any desired $s$, $v$, and $w$). In general, consider $N$ ``ground" states $\ket{g_1}, \dots, \ket{g_N}$ coupled to each other with $2 N - 2$ control fields via $N-1$ intermediate ``excited" states $\ket{e_1}, \dots, \ket{e_N}$ according to the Hamiltonian 
$H = \sum_{j=1}^N \ket{e_j}(\bra{g_j} \Omega_{j,j} + \bra{g_{j+1}} \Omega_{j,j+1})+ \textrm{h.c.}$ \cite{gorshkov13}. It is clear that this Hamiltonian supports a unique zero-energy dark state $\ket{D} = \sum_j c_j \ket{g_j}$ made up of ``ground" states alone, where the amplitudes $c_j$ are set by $c_j/c_{j+1} = - \Omega_{j,j+1}/\Omega_{j,j}$ and can be tuned to arbitrary values by tuning the ratios of the Rabi frequencies. In the case of $\ket{\u}$, the role of the 5 ``ground" states is played by $\ket{4}$, $\ket{5}$, $\ket{72 \textrm{S}_{1/2}, m_J = -1/2}$, $\ket{6}$, and $\ket{3}$, while the role of the ``excited" states is played by the 4 intermediate states.

We choose a DC electric field of 0.5 V/cm. At this field, the energy difference $\Delta E_\textrm{Stark}$ between $\ket{1}$ and $|70$P$_{3/2},m_J\!=\!-1/2\rangle$ is $\approx (2 \pi) 60$ MHz (while the energy difference between $\ket{4}$ and $\ket{5}$ is much larger $\sim (2 \pi)$ 220 MHz). At the same time, this DC electric field is sufficiently weak that the admixture of other states into bare states $\ket{1}$, $\ket{4}$, and $\ket{5}$ remains small ($\lesssim 0.2$), justifying the assumption of negligible induced dipole moments and the use (see below) of transition dipole moments corresponding to a vanishing electric field.  With $\Delta E_\textrm{Stark}/2\pi$ = 60 MHz, we have the following ladder of energy scales: $(\Delta E_\textrm{Stark}/2 \pi = 60 \textrm{ MHz}) \gg (\Omega_\textrm{dr}/2 \pi = 10 \textrm{ MHz}) \gg (\Omega_{\u,\d}/2 \pi = 2  \textrm{ MHz}) \gg (\omega_\textrm{EIT}/2 \pi = 300 \textrm{ kHz}) \gg (V_\textrm{dd}/2 \pi = 60 \textrm{ kHz}) \gg (\kappa/2 \pi = 10 \textrm{ kHz})$. This ladder ensures the following:
\begin{itemize}

\item
The condition $\Delta E_\textrm{Stark} \gg \Omega_\textrm{dr}$ (where $\Omega_\textrm{dr}$ is the Rabi frequency of the optical and microwave dressing fields) 
removes the degeneracy between Zeeman levels with different $|m_J|$. This enables the use of frequency selection for addressing desired transitions. For example, this condition ensures that the dressing lasers and microwaves do not Raman-couple $\ket{3}$ to $\ket{4}$, or $|70\textrm{P}_{3/2},m_J \!=\!-1/2\rangle$ to $|70\textrm{P}_{3/2},m_J \!=\!1/2\rangle$.  The DC Stark shift also allows us to avoid the two-photon resonant excitation of $\ket{69\textrm{D}_{3/2},m_J = 3/2, m_I \!=\! 1/2}$ (instead of $\ket{3}$) by $\Omega_\uparrow$ since $\ket{m_J = 3/2}$ moves out of two-photon resonance. 

\item
The condition $\Omega_\textrm{dr} \gg \Omega_{\u/\d}$ ensures that the control fields couple to the dark states $\ket{\u}$ and $\ket{\d}$ created by the dressing fields, but not to the bright states.  

\item
The condition $\Omega_{\u/\d} \gg \omega_\textrm{EIT}$ (where $\omega_\textrm{EIT} \sim \Omega_{\u,\d}^2/\Delta_{\u,\d}$ is the EIT linewidth \cite{fleischhauer05}) arises from the requirement $|\Delta_{\u}-\Delta_{\d}| \gg \Omega_{\d}$, which prevents two-photon resonant coupling of $\e_\u$ to $\Omega_\d$. Specifically, we take $\Delta_\u = - \Delta_\d = (2 \pi) 10$ MHz. Optical elements can then be used to ensure that $\e_\u$ and $\e_\d$ (whose frequencies $\omega_{1,\u}$ and $\omega_{1,\d}$ thus differ by $(2 \pi) 20$ MHz) are resonant with cavity modes at their respective polarizations. 

\item
The condition $\omega_\textrm{EIT} \gg V_\textrm{dd}$ ensures that the interactions are not strong enough to violate EIT and compromise dark-state polaritons. 

\item
Finally, the condition $V_\textrm{dd} \gg \kappa$ ensures that the Hamiltonian responsible for the fractional Chern insulator operates on an energy scale larger than the rate at which  photons leak out of the cavity. Decay rates at the 10 kHz level  
are reasonable 
 \cite{notcutt05}. 

\end{itemize}
It is worth pointing out that, at room temperature, the decay rates of the Rydberg states involved (69D, 70S, and 70P) are $\gamma_\textrm{R} \lesssim (2 \pi) 1$ kHz \cite{saffman10}, making these rates negligible compared to $V_\textrm{dd}$. 
It is also worth noting that the implementation of the purely atomic fractional Chern insulator via Eq.\ (\ref{eq:Seq}) requires a much simpler ladder of energy scales since the control Rabi frequency, the EIT linewidth, and cavity decay rate no longer enter:  $\Delta E_\textrm{Stark} \gg \Omega_\textrm{dr} \gg V_\textrm{dd} \gg \gamma_\textrm{R}$.

To derive Eq.\ (\ref{eq:Seq}) in the main text, we assumed that, for any atom $j$ on site $A$, the condition $[\sin(\omega_{1, \u} z_j/c)/\sin(\omega_{1, \d} z_j/c)] \exp[i (\omega_{2,\u}-\omega_{2,\d})z_j/c] \approx 1$ holds up to an $A$-dependent phase. We now verify this condition.  
On a given site, $z_j$ varies at most by the thickness of the atomic cloud, which needs to be smaller than the Rydberg blockade radius, $\approx 10$ $\mu$m, in order to ensure an intra-site excitation blockade. Hence, $\exp[i (\omega_{2,\u}-\omega_{2,\d}) z_j/c]$ will not vary appreciably with $j$ since $|\omega_{2,\u}-\omega_{2,\d}| \approx (2 \pi) 17 \textrm{ GHz}$, which is the energy separation between 70S and 69D.  
Now $|\omega_{1,\u} - \omega_{1,\d}| = |\Delta_\u - \Delta_\d| = (2 \pi) 20$ MHz, while $\omega_{1,\u} = (2\pi) 384$ THz (i.e.\ the D2 line in ${}^{87}$Rb). Thus, 
as long as the cavity length is less than $\approx 10$ cm, the two modes get $< 0.01$ out of phase with each other, ensuring that $[\sin(\omega_{1, \u} z_j/c)/\sin(\omega_{1, \d} z_j/c)] \approx 1$.

We now derive the dependence of $t_{AB}$ in Eq.\ (\ref{eq:final}) on the dressing parameters $s$, $v$, and $w$. Let $\mathbf{d}$ be the dipole-moment operator, so that $d^0 = d^z$ and $d^\pm = \mp (d^x \pm i d^y)/\sqrt{2}$ are the three components of the corresponding irreducible spherical tensor. 
Then, from the Wigner-Eckardt theorem, referring to Fig.\ \ref{fig:scheme}(b) in the main text, $-\bra{5}d^0\ket{1} = \bra{6}d^0\ket{2} = \mu_{26} = d \sqrt{3/5}$ and  $-\bra{4}d^+\ket{1} = \bra{3}d^-\ket{2} = \mu_{23} = d \sqrt{2/5}$ for some reduced matrix element $d$. Taking $R$ to be the distance between Rydberg atoms $i \in A$ and $j \in B$ and dividing by $1/(4 \pi \epsilon_0 R^3)$, the dipole-dipole Hamiltonian between  spins $i$ and $j$ becomes
\ba
 H_{ij} &=& (1 - 3 \cos^2 \theta) (d_i^0 d_j^0 + \tfrac{1}{2} (d_i^+ d_j^- + d_i^- d_j^+)) \nonumber \\
 &&- \tfrac{3}{2} \sin^2 \theta \left[e^{-2 i \phi} d_i^+ d_j^+ + \textrm{h.c.}\right], \\
&\rightarrow & (1 - 3 \cos^2 \theta) \Big[ \mu_{26}^2 (\ket{51}\bra{15}  +\ket{62}\bra{26} - \ket{61}\bra{25}
\nonumber \\
&&  - \ket{52}\bra{16})  - \tfrac{1}{2} \mu_{23}^2 (\ket{32}\bra{23} + \ket{41}\bra{14}) \Big] \label{eq2} \\
&& - \tfrac{3}{2} \sin^2 \theta \left[e^{-2 i \phi}  \mu_{23}^2 (\ket{42}\bra{13} + \ket{24}\bra{31})\right] + \textrm{h.c.} \nonumber \\
&\rightarrow& 
t_{ij} \sigma^i_{\u \d} \sigma^j_{\d \u} 
+ \textrm{h.c.},
\ea
where
\ba
t_{ij} &=& (1 - 3 \cos^2 \theta) ( \mu_{26}^2 w_i^* w_j  - \tfrac{1}{4} \mu_{23}^2 (s_i^* s_j + v_i^* v_j))  \nonumber \\ && \,\,\, + \frac{3}{4} \sin^2 \theta \mu_{23}^2 (e^{-2 i \phi} v^*_i s_j + e^{2 i \phi} v_j s_i^*)
\ea
and where the first ``$\rightarrow$" projects the Hamiltonian onto states $\ket{1}$ through $\ket{6}$, while the second ``$\rightarrow$" projects the Hamiltonian onto states $\ket{\u}$ and $\ket{\d}$. The first projection is dictated by energy conservation and relies on the fact that the strength of dipole-dipole interaction between two sites is smaller than the splitting between different $|m_J|$ introduced by the DC electric field \cite{gorshkov11b,gorshkov11c,gorshkov13}. The second projection is also dictated by energy conservation and relies on the fact that the auxiliary dressing fields used to define $\ket{\u}$ and $\ket{\d}$ split these two (dark) states from all the other (bright) states by an energy larger than  the strength of dipole-dipole interaction between two sites \cite{gorshkov11b,gorshkov11c,gorshkov13}. 

\begin{figure}[t!]
\begin{center}
\includegraphics[width = 0.35 \columnwidth]{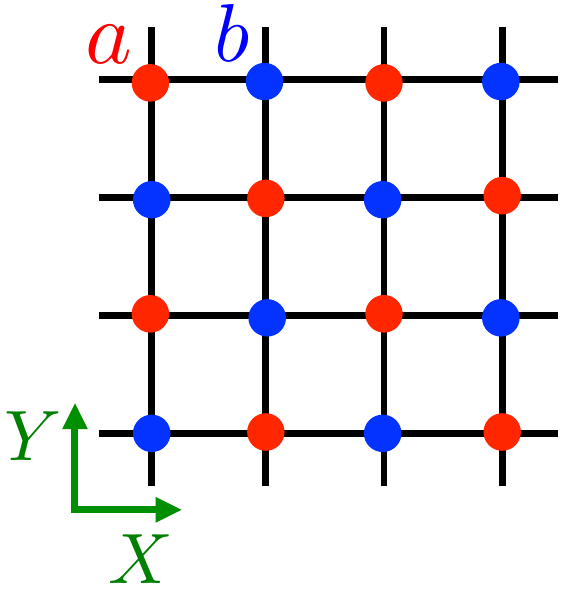}
\caption{Checkerboard lattice used to create the fractional Chern insulator. The parameters $s$, $v$, and $w$ used to define $\ket{\u}$ are different on the two sublattices ($a$ and $b$). In addition, the sign of $w_a$ and $w_b$ alternates every other row (not shown).\label{fig:lattice}}
\end{center}
\end{figure}
Finally, we present the specific  values of $s$, $v$, and $w$ that were used to obtain Fig.\ \ref{fig:FCI} in the main text. As in Ref.\ \cite{yao13c}, 
we consider a checker-board lattice, shown in Fig.\ \ref{fig:lattice}, consisting of an $a$ sublattice and a $b$ sublattice, so that parameters $s$, $v$, and $w$ are different on the two sublattices. Specifically, we parameterize $s_{a/b} = \sin(\alpha_{a/b}) \sin(\theta_{a/b})$, $v_{a/b}= \sin(\alpha_{a/b}) \cos(\theta_{a/b}) e^{i \phi_{a/b}}$, $w_{a/b} = \cos(\alpha_{a/b}) e^{i \gamma_{a/b}}$. For Fig.~\ref{fig:FCI} in the main text,
we chose $\{\theta_a, \theta_b, \phi_a, \phi_b, \alpha_a, \alpha_b, \gamma_a, \gamma_b\} =  \{0.87, 1.01, 2.79, 3.44, 2.37, 1.31, 4.71, 6.35\}$. We make a further modification by changing the sign of $w_{a}$ and $w_{b}$ every other row. Without increasing the size of the unit cell (2 sites), this modification plays an important role in allowing us to flatten the topological band. Finally, in Fig.\ \ref{fig:FCI} of the main text,
$\Theta_0 = 0.68$ and $\Phi_0 = 2.60$ are, respectively, the polar and the azimuthal angles of the DC electric field ($\hat z$) in the coordinate system determined by the $X$-$Y$ plane in which the square lattice is sitting.

\end{document}